\documentstyle[psfig,12pt,a4]{article}
\begin{document}%

\title{A geometrical approach
to non-adiabatic transitions in quantum theory:
applications to NMR, over-barrier reflection and 
parametric excitation of quantum oscillator.}
\author{
 M. S. Marinov\footnote{Prof. M. S. Marinov passed away on January 17, 2000.}
and E. Strahov\footnote{strahov@physics.technion.ac.il}\\
{\sl Physics Dept., Technion-Israel Institute of Technology}\\
{\it Haifa, 32000 Israel}
}
\date{21.03.2000}
\maketitle
   \begin{abstract}
This article deals with non-adiabatic processes (i.e. processes excluded
by the adiabatic theorem) from the geometrical (group-theoretical) point 
of view. An approximated formula for the 
probabilities of the  non-adiabatic transitions is derived in the 
adiabatic regime for the case when the parameter-dependent
Hamiltonian  represents a smooth curve in  the Lie algebra
and  the quantal dynamics is determined by the  corresponding
Lie  group evolution  operator.  
We treat the spin precession in a time-dependent magnetic field and the
over-barrier reflection problem in a uniform way using the first-order 
dynamical equations on $SU(2)$ and $SU(1.1)$ group manifolds correspondingly.
 A  comparison with  analytic solutions for simple solvable models is 
provided. 
     \end{abstract}
\section{Introduction.}
It is well known that probabilities of
transitions induced by a time dependence of the Hamiltonian are
suppressed if the dependence is slow. This statement 
formulated and proved by Born and Fock\cite{born}
 is presented in standard textbooks (e.g. by Messiah\cite{messiah})
and known as the {\em adiabatic theorem}.
The adiabatic theorem does not prove that the transitions are forbidden,
it means just that the probabilities are suppressed exponentially and
vanish to any finite order of the standard perturbation theory. 
Transitions of this type should not be discarded, however, if they result in
special phenomena, even though relatively rare ones.
The over-barrier reflection and the spin-flip in a time-dependent
 magnetic field 
slowly deviating from  the original direction may serve as  simple
examples of such phenomena. Usually, the
adiabatic character of the process suggests a way to evaluate its
probability, like in the quasi-classical approximation.

Corrections to the adiabatic theorem were considered in a number of
works, especially with application to physical systems 
with  two non-degenerate  energy levels. Dykhne\cite{dykhne}
considered a Hamiltonian $\hat H(t)$ with 
two eigenstates $\Psi_1(t), \Psi_2(t)$ 
which is analytic in time. He found the transition probability $p_{12}$ 
for a system, prepared at time $-\infty$ in the eigenstate $\Psi_1(-\infty)$
 to pass to the eigenstate $\Psi_2$ as $t$ runs from $-\infty$ to $+\infty$.
This probability is expressed by the formula:
\begin{equation}
p_{12}\sim\exp\Bigl(-2 T |{\rm Im} \int^{t_c}_{0} (E_2(t) - 
E_1(t))dt|\Bigr).
\end{equation}
Here $t_c$ is a point in the complex time plane in which $E_2(t)$ and 
$E_1(t)$ cross, and $T$ is a time-scale parameter (large in the adiabatic 
limit) over which $\hat H(t)$ changes essentially.
 A rigorous derivation
of Dykhne's result was given by Davis and Pechukas\cite{davis}.  
Suominen, Garraway and Stenholm\cite{suominen},\cite{stenholm} have 
applied the Dykhne, Davis and Pechukas approach to two-level solvable 
models. In particular, an adiabatic behavior of the Landau and Zener 
\cite{landau1}, \cite{zener} model was considered.    It appears that for 
this model 
the Dykhne, Davis and Pechukas method gives the exact answer.

From the works of Berry\cite{berry2}, Joye, Kunz and Pfister\cite{joye}, 
Jak\v si\'c and Segert\cite{jak1},\cite{jak2} it becomes clear that 
similar to Berry adiabatic phase\cite{berry1},
the transition 
probabilities induced by a time dependence of the Hamiltonian are 
connected with the geometry of the parameter space.
In particular, when the evolution of a system is described by the 
Hamiltonian of the form
\begin{equation}\label{ham}
\hat H(s)={\bf n}(s)\cdot\sigma, \;\; |{\bf n}(s)| =1,
\end{equation}
(where ${\bf n}(s)$ is a parameter dependent unit vector and 
$\sigma=(\sigma_1, \sigma_2, \sigma_3)$ are Pauli matrices) in the 
first-order adiabatic perturbation theory the transition (spin-flip) 
$W_{\pm}$ is determined by the Fourier transform \cite{berry2}, 
\cite{joye}, \cite{jak1}, \cite{jak2}:
\begin{equation}
W_{\pm} \sim |\int^{+\infty}_{-\infty}\exp(-2iTs)\chi (s)ds |^2
\end{equation}
\begin{equation}
\chi(s)= i/2 |n'(s)|\exp(-i\varsigma(s)),
\end{equation}
where $|{\bf n}'(s)|$ is related with the Riemannian length element 
$dl(s)=|{\bf n}'(s)|ds$ of the unit sphere. The function $\varsigma(s)$ is 
given by the integral
\begin{equation}
\varsigma(s)=\int^{s}_{0}\kappa_g(s)ds,
\end{equation}
and $\kappa_g(s)$ is the geodesic curvature of a path ${\bf n}(s)$:
\begin{equation}\label{curvature}
\kappa_g(s)=\frac{{\bf n''}(s)\cdot ({\bf n'}(s)\times {\bf n}(s))}
{|{\bf n'}(s)|^2}.
\end{equation}
We note that the Hamiltonian (\ref{ham}) defines a curve in the Lie 
algebra $su(2)$. The corresponding evolution operator belongs to the 
fundamental $(2\times 2)$ representation of the group $SU(2)$
and the transition probability appears to be completelly
determined by the geometric properties of the $SU(2)$
group homogeneous space $S^2=SU(2)/U(1)$ (which serves
as the parameter space for this particular situation).

It is the 
main purpose of this work to establish a relation between the probabilities 
of the non-adiabatic transitions and the geometry of the parameter 
space in a more general than the mentioned above case. Namely, we 
consider a situation when the commutator algebra of the Hamiltonian 
operators $\hat H(s)$ at different values of parameter $s$ is closed for 
all $s$, constituting an arbitrary Lie algebra ${\cal G}$. 
This condition makes it possible
to reformulate the original problem about the evolution
of a quantum state in terms of a first-order
dynamic equation on the group manifold (section 2). The adiabatic
solution of this equation  is constructed in section 3,
and the integral expression for  the Lie algebra element,
determining transition probabilities, is derived. In section 4
our approach is  applied to the spin  precession in
a variable external field. The adiabatic approximation
for the spin precession\cite{berry2},
\cite{joye}, \cite{jak1}, \cite{jak2} is reconstructed.
 Applied to the ordinary one-dimensional
Schr\"odinger equation, our method gives the well-known WKBJ
solution as the first adiabatic approximation.
In the leading (second-) order approximation 
our approach  leads to Bremmer's formula\cite{bremmer}
for the over-barrier reflection (section 5).
In section 6  
an expression  for the transition  probability due to a 
 parametric
excitation of a quantum oscillator is obtained from
our results.  Section 7
is devoted to a comparison of  our calculations with 
exact solutions for a number of analytically solvable models.
\section{Dynamic equation on group manifolds}
We consider the linear operator (matrix) equation of the form
   \begin{equation}\label{1}  
\partial\hat{G}/\partial t=\hat{B}(t)\hat{G},
\;\;\;\hat{G}(t_0)=\hat{I},
   \end{equation}
where $\hat{I}$ is the unit operator and $\hat{B}(t)$ has a given time
dependence. This is a pattern for a number of physical problems,
including the one-dimensional Schr\"{o}dinger equation and the spin
precession in a time-dependent magnetic field. If the commutator algebra
of operators $\hat{B}(t)$ is closed for all $t$, constituting a Lie
algebra $\cal G$, one deals actually with the first-order dynamical
equations on the Lie group $\sf G$, generated by $\cal G$. Now the
problem can be written  in terms of the Cartan -- Maurer one-form,
   \begin{equation}\label{CM}  
dg\;g^{-1}=b(t)dt,\;\;\;g(t)\in{\sf G},\;\;\;b\in{\cal G},
   \end{equation}
with the initial condition $g(0)=e$ - the unit element of ${\sf G}$.
Special problems are those where $b(t)$ belongs to a Cartan subalgebra of
$\cal G$, i.e. $b(t)\in{\cal H\subset G}, \forall t$. In a case like
that, Eq.(\ref{CM}) is integrated immediately,
   \begin{equation}\label{cartan} 
g(t)=\exp\left[\int^t_{t_0}\!b(\tau)d\tau\right]\;\;\in\;{\sf H},
   \end{equation}
where $\sf H$ is the corresponding Abelian subgroup of $\sf G$.
In general, (\ref{CM}) is a set of (non-linear) differential equations which
cannot be reduced to quadratures.
It is notable that the desired group element may be shifted by a properly
chosen amount $g_0(t)$, so the equation is rewritten in an equivalent
form, $g=g_0(t)g_1$,
   \begin{equation}\label{gaugetransform}
dg_1\;g_1^{-1}=b_1(t)dt,\;\;\;
b_1(t)dt=g_0^{-1}\left(bdt-dg_0\;g_0^{-1}\right)g_0\in{\cal G}.
   \end{equation}
Thus the problem may be reduced to a more tractable one.

Let us restrict ourselves to problems where $b(t)$ approaches a Cartan
subalgebra $\cal H$ asymptotically, as $t\rightarrow\pm\infty$, and
evaluate the transition probability between eigen-states of operators
representing $\cal H$. The $S$-operator given by the
following limit may be used (provided this limit exists):
\begin{equation}  
\hat{S}=\lim_{t,t_0\rightarrow\pm\infty}
\exp\left[-\int^t_0\hat{B}_+(\tau)d\tau\right]\hat{G}_{t_0}(t)
\exp\left[-\int^0_{t_0}\hat{B}_-(\tau)d\tau\right],
  \end{equation}
where $\hat{B}_\pm(t)\in{\cal H}$, and
$\lim_{t,t_0\rightarrow\pm\infty}\|\hat{B}(t)-\hat{B}_\pm(t)\|=0$.
The probability of transition between the states given by the density
operators $\hat{P}_\pm$ at $t\rightarrow\pm\infty$ is
   \begin{equation}\label{transition}  
W_\pm\equiv\lim_{t,t_0\rightarrow\pm\infty}{\rm Tr}
\left[\hat{P}_+\hat{G}(t)\hat{P}_-\hat{G}^\dagger(t)\right]
={\rm Tr}\left(\hat{P}_+\hat{S}\hat{P}_-\hat{S}^\dagger\right),
  \end{equation}
since we assume that $\hat{P}_\pm$ commute with $\hat{B}_\pm$.
\section{ The adiabatic approximation}
At any given time $t$, the driving force $b(t)\in{\cal G}$ may
be reduced to a Cartan subalgebra $\cal H$, and the group
element is decomposed respectively
   \begin{eqnarray}\label{7}  
& b(t)=v(t)\beta(t)v(t)^{-1},\;\;\;\beta(t)\in{\cal H},\\
& g(t)=w(t)h(t)w(t)^{-1},\;\;\;h(t)\in{\sf H}.
   \end{eqnarray}
Remarkably, if $v$ has no $t$-dependence, that would fix
the subalgebra $\cal H$ for all $t$, and $g$ would be obtained
immediately, like in Eq.(\ref{cartan}). We consider the problems where $b(t)$
belongs to the Cartan subalgebra asymptotically, at $t\rightarrow\pm\infty$,
so $\lim v(t)=e$. The equation resulting from (\ref{CM}) would be
   \begin{equation}\label{Eq.9}
w\left(dh\;h^{-1}+w^{-1}dw-hw^{-1}dwh^{-1}\right)w^{-1}=
v\beta v^{-1}dt.
  \end{equation}
Splitting this equation to the subalgebras $\cal H$ and $\cal G\setminus
H$, we get a set of differential equations for $h$ and $w$.
We find an approximate solution of the Eq.(\ref{Eq.9})
for {\em adiabatic} processes, where the $t$-dependence of $v$ is slow,
the derivative $dv/dt$ is small, 
and the condition
\begin{equation}\label{Eq.10}
\Vert v^{-1}dv/dt\Vert\ll\Vert\beta\Vert
\end{equation}
is satisfied. Here the norm $\Vert y\Vert$ for an arbitrary
element $y$ of the Lie algebra ${\cal G}$ is introduced,
\begin{equation}\label{Eq.11}
\Vert y\Vert=\sqrt{{\rm Tr}(YY^\dag)},\;\;\;
y\in{\cal G} 
\end{equation}
and $Y$ is the matrix belonging to the adjoint  representation
of the Lie algebra ${\cal G}$ and corresponding to the
Lie algebra element $y$. When the condition (\ref{Eq.10}) holds, 
$w$ is always close to $v$, and the
deviation of $g(\infty)$ from the subgroup  $\cal H$ is negligible.
This is the meaning of the adiabatic theorem: the eigen-states of
operators belonging to the subalgebra $\cal H$ are not subject to
transitions.

The unknown group element may be  replaced by $w=v\exp(-\omega)$, and it
is assumed that $\omega\in{\cal G\setminus H}$. Small deviations from the
adiabatic limit, producing non-adiabatic transitions, are obtained
if we consider the first approximation in $\omega$, which is expected
to be of the order of $v^{-1}dv$, discarding all higher-order terms.
The result is
\begin{equation}  
dh\;h^{-1}-R(h)(v^{-1}dv-d\omega)=[\beta+(\omega\beta-\beta\omega)]dt,
  \end{equation}
where $R(h)\eta\equiv h\eta h^{-1}-\eta, \forall\eta\in{\cal G}$
(note that $R(h)\eta=0$, if $\eta\in{\cal H}$).
Separating the zero- and the first-order terms, we get two equations
   \begin{eqnarray}  
dh_0\;h_0^{-1}=\beta(t)dt,\;\;\;{\rm so}\;\;\;
h_0=\exp\left[\int^t_{t_0}\beta(\tau)d\tau\right]\in{\sf H},\\
R(h_0)\partial\omega/\partial t+[h_{0}^{-1}\partial h_{0}/\partial t,\omega]=
R(h_0)(v^{-1}\partial v/\partial t).
   \end{eqnarray}
The latter equation is also integrated immediately,
   \begin{equation}  
R(h_0^{-1})\omega=\int_{t_0}^tR(h_0^{-1})(v^{-1}\dot{v})d\tau,\;\;\;
\dot{v}\equiv\partial v/\partial t|_{t=\tau}.
    \end{equation}
In the asymptotics, as soon as $t, t_{0}\rightarrow\pm\infty$,
 we get
   \begin{equation}\label{14}  
\gamma\equiv\lim_{t, t_{0}\rightarrow\pm\infty}
R(h_0^{-1})\omega=
\int_{-\infty}^\infty R(h_0^{-1})(v^{-1}\dot{v})d\tau,
   \end{equation}
and this element of $\cal G$ determines the transition probability
amplitude. In order to see that, let us insert the asymptotic value of
the operator representing the group element
   \begin{equation}  
g=v\exp(-\omega)h\exp(\omega)v^{-1}\approx (e-\omega)h_0(e+\omega)
\approx h_{0}(e-R(h_0^{-1})\omega)  
 \end{equation}
in Eq.(\ref{transition}) for the transition probability. 
Assuming that $\hat{P}_+$
and $\hat{P}_-$ represent different (orthogonal) eigen-states of
operators corresponding to $\cal H$, so 
that $\hat{P}_+\hat{P}_-=0=\hat{P}_-\hat{P}_+$, one gets
the following expression for the transition probability
in the leading (second-) order
   \begin{equation}\label{prob}
W_\pm={\rm Tr}(\hat{P}_+\hat{\Gamma}\hat{P}_-\hat{\Gamma}),
   \end{equation}
where $\hat{\Gamma}$ is the operator representing $\gamma\in{\cal G}$
in Eq.(\ref{14}).
Note  that any value may  be taken for $t_{0}$ in Eq.(19);
changing it, say, to $t'_0$, would  result
in  a constant gauge substitution of $\gamma$ for $\gamma '$,
\begin{equation}
\gamma '=(h_{0}')^{-1}\gamma h_{0}',\;\;\; 
h_{0}'=\exp\left[\int_{t_0'}^{t_0}\beta(\tau)d\tau\right].
\end{equation}
That would not change the probability in (\ref{prob}).
The convergence of the integral in (\ref{14}) depends on how fast the driving
force $b(t)$ is approaching its asymptotics in $\cal H$. It is noteworthy
that the present result extends the standard perturbation theory. If
$b(t)=\beta_0+\lambda b_1(t)$, where $b_1(t)\rightarrow 0$ at $\pm\infty$,
then to the first order in $\lambda$ one has to set
$h_0(\tau)=\exp(\beta_0\tau)$ in (\ref{14}), and the result is
 an extension of
the Born approximation. In general, $\gamma$ indicates the deviation from
the adiabatical limit. Even if the perturbation is not small absolutely,
$\gamma$ may be small because of two different reasons: i) the change of
$v$ is slow, though it may be not close to unity, which is the case for
small perturbations, ii) the deviation of $b(t)$ from $\cal H$ takes place
during a small time interval, and the integral is small as a result of that.
\section{Spin-flip in a variable magnetic field}
  The spin precession in a time-dependent magnetic field, the fundamental problem 
for NMR\cite{abragam}, is determined by the Bloch equation 
for the spinor wave function,
\begin{equation}\label{eq18}
id\psi / dt =({\cal B} \cdot \sigma)\psi ,
\end{equation}
where ${\cal B}\equiv \mu {\bf B}(t)$, ${\bf B}(t)$ is a variable
magnetic field vector, $\mu$ is the particle magnetic moment, and
$\sigma$ are the Pauli matrices. The fundamental solution of
 Eq.(\ref{eq18}) is given by a unitary $2\times 2$ matrix, so the 
group is $SU(2)$. The matrix $\hat G$ is given by Eq.(\ref{1})
 with $\hat B = -i({\cal B}\cdot \sigma)$.

Let us consider, for instance, the case where a pulse is applied in the 
$x$-direction, while the $z$-component is constant, 
${\bf B}=\{B_1(t), 0, B_0\}$, and $B_1(t)\rightarrow 0$ as 
$\rightarrow \pm\infty$. The adiabatic approximation holds if $B_1(t)$
is a slow function of $t$. The elements which appear in (\ref{7}) are
\begin{equation}
\beta(t)=2\mu[B^2_0 + B^2_1(t)]^{1/2}J_3, \;\;\; v(t)=\exp(\theta J_2),
\end{equation}
where tan $\theta = -B_1(t)/B_0$, and $J_{a}$ is the basis in ${\cal G}$,
represented by $\frac{i}{2}\sigma_{a}$. 
For this particular representation , we find from the Eq.(\ref{Eq.11}) ,
that
\begin{equation}\label{Eq.19}
\Vert v^{-1}dv/dt\Vert=\vert d\theta/dt\vert ;\;\;\; 
\Vert\beta\Vert=\mu[B^2_0+B^2_1(t)]^{1/2}
\end{equation}
and the general condition of applicability (\ref{Eq.10}) leads to
\begin{equation}\label{spinapl}
\vert d\theta/dt\vert\ll\mu[B^2_0+B^2_1(t)]^{1/2}
\end{equation}
The spin-flip, as given by 
Eq.(\ref{14}), is determined by $\gamma=A_{+-}J_2$, and 
\begin{equation}\label{spinamplitude}
A_{+-}=\int^{\infty}_{-\infty} e^{2i\alpha(\tau)}\dot\theta d\tau, \;\;\;
\alpha(\tau)=\mu\int^{\tau}_{\tau_1}[B^2_0 + B^2_1(t)]^{1/2} dt.
\end{equation}
The spin probability is $W_{+-}=|A_{+-}|^2$. The accuracy of the 
approximation has been checked for a field where the exact analytical 
solution is available (see section 7).

In the formula (\ref{spinamplitude}) $dl(\tau)=\dot\theta d\tau$
is the Riemannian length element on the path
${\bf n}(\tau)=(\sin\theta,0,\cos\theta)$. The geodesic curvature
determined by Eq.(\ref{curvature}), is equal to zero.
In a more general case of the magnetic field configuration
${\bf B}=\vert B\vert(\sin\theta\cos\phi,\sin\theta\sin\phi,cos\theta)$
the group element $g_0$ leading to Eq.(\ref{gaugetransform})
will be chosen as $g_0=\exp(-2\phi J_3)$. This transformation
alters the phase $\alpha(\tau)$ in Eq.(\ref{spinamplitude}) 
to
$\;\; \mu\int^{\tau}_{\tau_1}[\sin^2\theta+(\cos\theta-\frac{\dot\phi}{\vert 
B\vert})^2]^{1/2}\vert B\vert dt$. An expansion of $\alpha(\tau)$
to the lowest-order non-vanishing in $\frac{\dot\phi}{\vert B\vert}$
leads to 
Berry\cite{berry2}, Joye, Kunz and Pfister\cite{joye},
Jak\v si\'c and Segert\cite{jak1},\cite{jak2}
result for the spin-flip probability.
\section{Over-barrier reflection}\label{reflection}
The Schr\"odinger equation, $\Psi''-U(x)\Psi=-k^2\Psi$ is
equivalent to the following first-order problem for the two-component
function $\psi(x)$, satisfying the equation
$$\frac{d\psi}{dx}=\hat{B}(x)\psi,$$
where
\begin{equation}   
\psi=\left(\begin{array}{c}
            \Psi'-ik\Psi\\
            \Psi'+ik\Psi \end{array}\right),\;\;\;
\hat{B}(x)\equiv-i\left(\begin{array}{cc}
            k-U/2k & U/2k \\
            -U/2k & -k+U/2k \end{array}\right).
    \end{equation}
Thus the coordinate $x$ plays the role of the time parameter $t$.
For the plane wave moving in the positive direction, $\Psi=C\exp(ikx)$,
so the upper component of $\psi$ vanishes, and reflection is like the
spin flip. The problem of barrier penetration is represented by 
Eq.(\ref{1}), where $\hat{B}^\dagger=-\sigma\hat{B}\sigma$
($\sigma\equiv\sigma_3$ is the diagonal Pauli matrix). Thus
$\hat{G}^\dagger=\sigma\hat{G}^{-1}\sigma$, the probability current
$j\equiv-\frac{1}{2k}\bar{\psi}\sigma\psi=
-\frac{i}{2}(\Psi'\bar{\Psi}-\Psi\bar{\Psi}')$ is
conserved, so we are dealing with the two-dimensional representation of
the group ${\sf G}=SU(1,1)$.
For $U(x)$ decreasing rapidly as
$x\rightarrow\pm\infty$, one has the transfer matrix
  \begin{equation}
\hat{T}\equiv \lim_{x,x_0\rightarrow\pm\infty}\hat{G}(x)=
\left(\begin{array}{cc}
            a & \bar{b}\\
            b & \bar{a}\end{array}\right), \;\;\;|a|^2-|b|^2=1.
 \end{equation}
The penetration probability amplitude is $1/|a|^2$ and the reflection
probability is $|b/a|^2$.

As soon as $\det\hat{B}=k^2-U(x)\equiv p^2(x)$, the Abelian subgroup is
${\sf H}=U(1)$ in the region where $p^2>0$, and $\sf H=R$ under the barrier,
where $p^2<0$. The $2\times 2$ matrix diagonalizing $\hat{B}$ is
\begin{equation}  
\hat{V}=\left(\begin{array}{cc}
\cosh\eta & \sinh\eta\\
\sinh\eta & \cosh\eta
\end{array} \right),\;\;\;\;\exp(2\eta)\equiv\frac{p}{k}.
    \end{equation}

Being applied to  the plane wave , moving in  the positive 
direction, the group element in the  first-order adiabatic  approximation
 $g\approx vh_0v^{-1}$ leads to the following
expression for the wave function:
\begin{equation}
\Psi(x) = C\left[\cosh\eta(x)e^{-\eta(x)
            +i\int^x_{x_0}p(\xi)d\xi}+
          \sinh\eta(x)e^{-\eta(x)-i\int^x_{x_0}p(\xi)d\xi}
            \right]
\end{equation}
In the limit $x\rightarrow +\infty$ the parameter
$\eta(x)$ goes to zero, 
\begin{equation}
\lim_{x\rightarrow +\infty}\Psi(x)=
C\exp\left(i\int_{x_0}^{x}p(\xi)d\xi\right)
\end{equation}
and the over-barrier reflection is absent in the first-order
adiabatic  approximation.
Note  that in  the framework of  this
 approximation the elements of  the matrix $\hat{V}$
are considered to  be slowly dependent on $x$, and $\exp(-\eta(x))=
\sqrt{\frac{k}{p(x)}}$. Thus a familiar WKBJ expression
for  the wave function:
\begin{equation}\label{WKBJ}
\Psi(x)=C_1\sqrt{\frac{k}{p}}\exp\left( i\int^x_{x_0} p(\xi)d\xi\right)
       +C_2\sqrt{\frac{k}{p}}\exp\left(-i\int^x_{x_0} p(\xi)d\xi\right).
  \end{equation}
is reconstructed.
Remarkably, $v$ belongs to a one-parameter subgroup of $SU(1,1)$, which
makes the calculations simpler than in the general problem of spin 
precession
(cf. section 4).
The over-barrier reflection  is determined by (\ref{14}), 
where the  element $v$  
is  represented  by  the $2\times 2$ matrix $\hat V$ , and  $h_{0}$ -by  the
matrix:
\begin{equation}
\hat{H_0}=\left(\begin{array}{cc}
e^{-i\int_{x_0}^{x}pdx} & 0\\
0 & e^{i\int_{x_0}^{x}pdx}
\end{array} \right)
\end{equation}
The  probability  of  the  over-barrier  reflection is $R=|A|^2$,
where
\begin{equation}\label{amplituda}
A=\frac{1}{4}\int_{-\infty}^{+\infty}e^{2i\int_{x_0}^{x}pdx}
\frac{U'(x)}{k^2-U(x)}dx
\end{equation}
The proposed method is valid when the unequality 
(\ref{Eq.10}) is satisfied. This leads to  the same condition
of applicability as in the WKBJ approximation,
\begin{equation}\label{wkbcondition}
|dp(x)/dx|\ll p^2(x)
\end{equation} 

Eq.(\ref{amplituda}) coinsides with the over-barrier reflection
amplitude obtained by Bremmer\cite{bremmer}. Bremmer's approximation
was to divide a smooth potential into a large number of small
layers. The momentum $p(x)$ , being different in different layers,
was assumed to be constant throughout a range of a particular
layer. The ordinary  WKBJ solution Eq.(\ref{WKBJ}) was then
obtained by discarding all reflections  of the incident
wave at any layer's boundary. Assuming that only single reflections
at all boundaries of the layers take place, Bremmer found the 
over-barrier reflection amplitude.

An advantage of our procedure is that leading to the same result 
(Eq.(\ref{amplituda})) as 
Bremmer's approach, our derivation does not demand any assumptions
about  qualitative character of the wave reflection. Thus our method
shows that the usual condition of the validity of WKBJ approximation
(Eq.(\ref{WKBJ}))
is only needed in order  to get Bremmer's formula.  

The reflection amplitude (\ref{amplituda}) may be compared with the 
one given by Maitra and Heller\cite{maitra}. These authors use a
perturbative approach with the WKBJ states as the unpertubed basis.
According to Maitra and Heller the approximated reflection
amplitude $A_{M.H.}$ is given by the matrix element of an effective
potential between the usual WKBJ wave functions, i.e.
\begin{equation}
A_{M.H.}=\int_{-\infty}^{+\infty}U_{eff}(x,k)\frac{e^{2i\int^x 
p(y)dy}}{p(x)}dx
\end{equation} 
 The effective potential of Maitra and Heller is given by the formula:
\begin{equation}
U_{eff}(x, k) = \frac{-3(p'(x))^2}{4p^2(x)} + \frac{p''(x)}{2p(x)}
\end{equation}
Once Maitra and Heller use the perturbative arguments their 
expression for the reflection amplitude should be valid when the 
effective potential is small, i.e. 
\begin{equation}\label{maitra}
|U_{eff}(x, k)| \ll k^2.
\end{equation}
Comparing the above conditions with that of applicability of our 
approximation we can see that our approach has a wider range of validity
since  inequality (\ref{maitra}) follows from Eq.(\ref{WKBJ}).
 
 In the  limit  of $k^2\gg U(x)$
the momentum $p(x)$ becomes approximately equal to $k$.
Neglecting  $U(x)$ in comparison
 with $k^2$ in the integral (\ref{amplituda}),
and integrating by parts,
 we  obtain  the result
which  corresponds to perturbation  theory:
\begin{equation}
A_{pert}=\frac{1}{2ik}\int_{-\infty}^{+\infty}e^{2ikx}U(x)dx
\end{equation}         
\section{Parametric excitation of a quantum oscillator.}
The parametric excitation of a quantum oscillator is the  excitation of the 
oscillator under change of its parameters $m=m(t)$ and $\Omega=\Omega(t)$.
The general case with a time-dependent $m(t)$ and $\Omega(t)$ may be easily
reduced to $m$=const by  changing  variables $ t^{'}=\int\frac{dt}{m(t)}$,
$\Omega^{'}=m\Omega$.
The Schr\"odinger equation for the
wave function of the quantum oscillator
has the following form:
\begin{equation}
i\frac{\partial\psi}{\partial t}=
-\frac{\partial^2\psi}{\partial x^2}+\frac{1}{2}\Omega^2(t)x^2\psi
\end{equation}
For $\Omega(t)$ the asymptotic conditions 
\begin{equation}
\Omega(t)\longrightarrow \Omega_{\pm},\;\;
t\longrightarrow\pm\infty
\end{equation}
are assumed. The assymptotic stationary states are:
\begin{equation}
\phi_n^{\pm}(x,t)=\phi_n(x,\Omega_{\pm})e^{-i(n+\frac{1}{2})\Omega_{\pm}t},
\end{equation}
\begin{equation}
\phi_n(x,\Omega_{\pm})=
\left(\frac{1}{2^nn!}\sqrt{\frac{\Omega_{\pm}}{\pi}}\right)^{\frac{1}{2}}
\exp{\left(-\frac{\Omega_{\pm} x^2}{2}\right)}H_{n}(\sqrt{\Omega_{\pm}}x)
\end{equation}
The time-dependence of the quantum oscillator parameters
$m(t)$  and $\Omega(t)$ allows for
the  transitions between different stationary states.

A typical problem  is  to calculate the  probability
of transitions  
 $W_{mn}$ from the 
 state $\psi_n$ with  the asymptotic $\phi_{n}^{(-)}(x,t)$
at $t\rightarrow -\infty$ 
to the asymptotic state $\phi_{m}^{(+)}(x,t)$
at $t\rightarrow +\infty$.
As is well known (e.g. Baz', Zeldovich and Perelomov\cite{baz}), 
in order to determine this probability 
of transitions $W_{mn}$ it is sufficient to calculate the 
quantum mechanical coefficient $\theta$ of the 
over-barrier reflection from the one-dimensional potential
 of a particle with momentum
$p(x)= \sqrt{k^2 - U(x)} = \Omega(x)$, 
where $\Omega(x)$ is the frequency function of
the quantum oscillator.
However, the analytic solution of the over-barrier reflection 
problem is known only for a number of special cases. 
When it is impossible to find an analytic solution for
the problem of the over-barrier reflection, our approximated 
approach developed in section (\ref{reflection}) may be applied. 
The approximated expression for the
quntum mechanical coefficient $\theta$ is given by the formula
(\ref{amplituda}), i.e.:
\begin{equation}
\vartheta = \frac{1}{4}\left| \int^{+\infty}_{-\infty}
e^{2i\int^{t}_{t_0}\Omega(t)dt}\quad\frac{\Omega '(t)dt}{\Omega (t)}
\right|^{2}
\end{equation}  
When the parameter $\theta$ is determined, the probability of
transitions $W_{mn}$ may be calculated using the 
Perelomov and Popov\cite{perelomov} formula: 
\begin{equation}
W_{mn}=\frac{n_{<}!}{n_{>}!}\left|\sqrt{ 1-\vartheta}
 P^{\frac{\mid m-n\mid }{2}}_{\frac{\mid m+n\mid }{2}}
(\sqrt{1-\vartheta})\right|^2 ,
\end{equation}
where $n_{<} = min(m,n)$, $n_{>} = max(m,n)$ and $P^{m}_{n}(x)$ are the 
associated Legendre functions.
\section{Comparison with  exact solutions.}
\subsection{Spin precession
in a magnetic  field.}
The exact solution of  the spin precession 
problem of Section 4 is known\cite{rosen} for the magnetic field
\begin{equation}
{\bf B(t)}=\frac{1}{T}\left(\frac{\beta_1}{\cosh(\frac{t}{T})},0,\beta_0\right)
\end{equation}
Here $\beta_0/T$ is the asymptotical precession frequency, $T$ is
the pulse duration.
The applicability condition Eq.(\ref{spinapl}) allows for the application
of our approach, when the inequality
\begin{equation}\label{spininequality}
\beta_1/\beta_0\ll\sqrt{\beta_1^2+\beta_0^2}
\end{equation}
is satisfied. When $\beta_0>1$, the
inequality (\ref{spininequality}) holds
for all $\beta_1$
 and the process would  be adiabatic.
Respectively, the perturbation theory can be applied for 
$\beta_1\ll\beta_0$. As known
from the analytical solution of Eq.(\ref{1}) given in terms of the
hyper-geometric function,
\begin{equation}
W_{+-}=[\sin(\pi\beta_1)/\cosh(\pi\beta_0)]^2.
\end{equation} 
Calculating the integrals in (\ref{spinamplitude}), it is usefull to change
the variables as follows
\begin{equation}
\beta_1/\beta_0=\tan k,\;\;
\cos k\sinh(t/T)=\sinh\xi.
\end{equation}
Taking the integral for $\alpha$ with $\tau_1=0$, we have
\begin{equation}
\alpha(\xi)=\beta_0\xi+\beta_1\arctan(\tan k\tanh\xi),
\end{equation}
\begin{equation}
A_{+-}=\sin k\int_{0}^{\infty}\sin[2\alpha(\xi)]
\frac{\tanh\xi d\xi}{(\cosh^2\xi-
\sin^2k)^{\frac{1}{2}}}
\end{equation}
The latter integral is reduced to a real form,
as the pulse is symmetrical under the time inversion, so $\dot\theta$
is odd. The numerical calculation shows  a wonderful accuracy
of the approximation, namely,
\begin{equation}
A_{+-}\approx\frac{\sin(\pi\beta_1)}{\cosh(\pi\beta_0)}
\end{equation}
even for moderate values of $\beta_0$ , in a wide range of $\beta_1$
(see Figure 1).

\subsection{Over-barrier reflection for  the  potential 
$U=\frac{U_0}{1+e^{-\gamma x}}$.}
The  analytic  expression  for  the  reflection
amplitude  is
\begin{equation}
A=\frac{\sinh(\pi
\alpha(1-\sqrt{1-\beta}))}{\sinh(\pi\alpha(1+\sqrt{1-\beta}))}
\end{equation}
In the  above formula the parameters $\alpha=k/\gamma$
and $\beta=U_0/k^2$ were  introduced.
Perturbation theory  may  be  applied when  $k\gg U_0$, i.e.
$0<\beta\ll 1$. In  that case  the  reflection  probability  is  equal
to
\begin{equation}
\rho\simeq\frac{\pi\alpha^2\beta^2}{4\sinh^2(2\pi\alpha)}
\end{equation}
 Next, we consider  the  situation  when $k^2\geq U(x)$.
As follows from the inequality (\ref{wkbcondition}),
in the cases when
\begin{equation}
\beta/\alpha\ll1-\beta
\end{equation}
the over-barrier probability amplitude may be calculated
by our method.
It  is  suitable  to  change  the  variables $z=e^{\gamma x}$
and  calculate  the  integral  in  
the  exponent  of   formula (\ref{amplituda}).
We  obtain  the  following integral expression
for the over-barrier reflection probability amplitude:
\begin{eqnarray}
 A=
\frac{\beta}{4}
\int^{+\infty}_{0}
\frac{z^{2i\alpha-1}
\left(2\sqrt{1-\beta}
\sqrt{(z+1)((1-\beta)z+1)}+
2(1-\beta)z+
2-\beta\right)^{2i\alpha\sqrt{1-\beta}}dz}{\left(2\sqrt{(z+1)((1-\beta)z+1)}
+(2-\beta)z+2\right)^{2i\alpha}(1+z)((1-\beta)z+1)}
\end{eqnarray}
A comparison of  the  exact and approximate
(see Figure 2) probability amplitudes demonstrates a
very good  accuracy of  the approximation
(\ref{amplituda}), namely
\begin{equation}
\vert A\vert\simeq
\frac{\sinh\pi\alpha(1-\sqrt{1-\beta})}{\sinh\pi\alpha(1-\sqrt{1+\beta})}
\end{equation}
\section{Conclusions}
In  this  paper we  have  proposed an adiabatic  
approach to the   calculation  of probabilities for   
quantum transitions.  In  the case when the  one-parameter dependent
Hamiltonian represents a  smooth curve in a Lie  algebra,
the original Schr\"odinger equation was  interpreted
as the dynamical equation on the corresponding
group manifold. The main result  of  this work is expressed  by   
Eq.(\ref{14})  that determines the  Lie algebra element
responsible for the non-adiabatic transitions.

The problem of over-barrier reflection in one-dimensional
quantum mechanics is  very similar in our approach   to  the problem 
of  spin-flip
in a variable magnetic field (the difference lies in the
fact that  for the over-barrier reflection problem
the  introduced evolution operator is an element
of  the group SU(1.1), and not of SU(2), as the  spin evolution
operator).  

We have tested our approach on simple problems for which approximate 
solutions are known. In the case of a spin in a time-dependent magnetic 
field our procedure leads to a spin-flip amplitude Eq.(\ref{spinamplitude}).
In the adiabatic limit Eq.(\ref{spinamplitude}) would coincide with the 
 Berry\cite{berry2}, Joye, Kunz and Pfister\cite{joye},
Jak\v si\'c and Segert\cite{jak1},\cite{jak2} result for the spin-flip 
probability amplitude. The application of our procedure to the 
over-barrier reflection gives Bremmer's formula\cite{bremmer} in the 
leading (second-) order approximation. It is remarkable that in order to 
obtain Bremmer's result, the usual condition of the validity of WKBJ  
approximation (Eq.(\ref{WKBJ})) is only needed.

Being checked for  two  solvable models
(the spin-flip in  the Rosen-Zener  magnetic field and the over-barrier
reflection for  the potential $U=U_0/(1+e^{-\gamma x})$ ),
our adiabatic approximation not only gives
the exponentially small character of the probabilities of  the
non-adiabatic  processes, but completely
describes  the  qualitative behavior  of  these 
probabilities as  functions  of the external parameters.
The  integrals (\ref{spinamplitude}), (\ref{amplituda}) show the same  
behavior under  variation of  the magnetic field  amplitude
(the  amplitude of  the potential) as the  exact  solutions.
It  is  interesting to  note  that  in  spite
of  the   same  condition of  applicability as the WKBJ
approximation, our  approach is  very
successful in  the calculation of the  over-barrier
reflection while the usual WKBJ approximation gives 
zero answer in all orders. The  reason is  that the 
WKBJ approximation is  an  asymptotic series that
is unable  to take the  exponentially 
small variables  into account. 
\subsection*{Acknowledgements.}
For valuable comments and discussions that have contributed
to  this work , many thanks to  J. Avron, B. Block,
N. Krauss , D. Owen and  B. Segev. 
 
\end{document}